\documentclass[a4paper]{article}
\usepackage{amsfonts}
\usepackage{amssymb}
\usepackage[margin=2cm]{geometry}
\usepackage[dvips]{graphicx}

\usepackage[T1]{fontenc}
\usepackage[latin2]{inputenc}

\usepackage{amsmath}
\usepackage{bbm}

\begin{document}

%\draft
%\twocolumn

\title{Comment on ``Nanoscale Heat Engine Beyond the Carnot Limit''}

\author{Robert Alicki \\ 
  {\small
Institute of Theoretical Physics and Astrophysics, University
of Gda\'nsk,  Wita Stwosza 57, PL 80-952 Gda\'nsk, Poland}\\
}

\date{\today}
% \date{July 26, 2003}
\maketitle

In a recent Letter Rossnagel \emph{et.al.} discussed the model of a nanoscale engine in a form of a driven quantum harmonic oscillator coupled to two baths. The cold baths was an equilibrium system at the temperature $T_1$ while the hot bath was assumed to be a ``squeezed thermal bath'' characterized by two parameters: the ``temperature'' $T_2 > T_1$ and the squeezing parameter $r \geq 0$. The authors proved that the engine efficiency $\eta$ exceeds the Carnot bound  $\eta_C =  1-\frac{T_1}{T_2}$,
and for $r\to\infty$ the efficiency at maximal power $\eta \to 1$. This result does not imply violation of the Second Law of Thermodynamics, because the hot bath is in a nonequilibrium state and therefore, the meaning of the ``temperature'' is not clear. 
\par
However, there exists a way to define a canonical  frequency-dependent ``local temperature''  for any nonequilibirum stationary bath  and for a given bath's observable which determines the interaction of a ``thermometer'' with a bath. For example, consider a thermometer which is a quantum harmonic oscillator or two-level system with the frequency $\omega$ weakly coupled to a stationary bath by means of the interaction Hamiltonian  $H_{int} = (a + a^+)\otimes B$. Here $ a, a^+$ are bosonic or fermionic annihilation and creation operators and $B$ is the bath observable. One can easily derive the proper Master Equation for the density matrix of the thermometer (oscillator or spin) and show that the bath drives the thermometer to the thermal equilibrium state at the frequency-dependent temperature $T[\omega]$ determined by the following relation involving spectral density of the bath
\begin{equation}
e^{-\frac{\hbar\omega}{k_B T[\omega]}} = \frac{G(-\omega)}{G(\omega)} , \quad  G(\omega) = \int_{-\infty}^{+\infty} e^{i\omega t}\langle B(t) B\rangle_{bath} dt.
\label{ham_TLSQB}
\end{equation}
In the case of equilibrium bath the relation \eqref{ham_TLSQB} is satisfied with a fixed $T$ and for arbitrary $B$ (Kubo-Martin-Schwinger condition). 
Consider the bath used in \cite{1} which consists of many harmonic oscillators (modes) with quasi-continuous frequency spectrum  $\{\omega_j\}$ and each of them being in a squeezed state characterized by a fixed ``temperature'' $T_2$ and a squeezing parameters ${r}$. Strictly speaking such a state is not stationary, but can be easily made a such by ergodic averaging which does not change the mode populations. In this case the temperature $T[\omega]$ measured  by  a harmonic or spin thermometer  linearly coupled to the oscillator bath is given by the formula 
\begin{equation}
e^{-\frac{\hbar\omega}{k_B T[\omega]}} = \frac{\langle n\rangle +(2{\langle n\rangle +1) \sinh^2(r)}}{1+ \langle n\rangle +(2{\langle n\rangle +1) \sinh^2 (r)}}
\label{kms}
\end{equation}
where $\langle n\rangle= \frac{1}{e^{\frac{\hbar\omega}{k_B T_2}} -1}$. For $r =0$ we obtain the equilibrium value  $T[\omega] = T_2$ while for $r\to\infty$ , $T[\omega]\to \infty$, exponentially with $r$.
Moreover, using a slightly generalized formalism \cite{2}, \cite{3} for periodically driven open systems interacting with many baths, one can derive the entropy balance equation
\begin{equation}
\frac{d S}{dt} - \sum_{\{\omega\}} \frac{J_{\omega}}{T[\omega]} \geq  0
\label{balance}
\end{equation}
where $S$ is von Neumann entropy of the open system, $\{J_{\omega}\}$ are properly defined heat currents and the sum is taken over all relevant ``generalized Bohr frequencies'' which appear in the Master Equation. The equation \eqref{balance} is valid for a general stationary nonequilibrium bath, satisfies the Second Law and allows to derive the Carnot bound
with the temperatures $T_1$ and $T_2$ given by the averaged local temperatures corresponding to outgoing (negative) and incoming (positive) heat currents, respectively.
\par
Summarizing, there exists a consistent thermodynamical formalism with canonical definition of local temperatures which allows to extend the Carnot bound also to non-equilibrium baths, like those discussed in the Letter \cite{1}.

\end{document}